\documentclass[aps,prl,twocolumn,groupedaddress,amsmath,amssymb,nobalancelastpage]{revtex4-1}
\usepackage{graphicx}  
\usepackage{bm}        

\begin{document}

\title{Disorder Protected and Induced Local Zero-Modes in Longer-Range Kitaev Chains}
\author{Simon Lieu}
\email{simonklieu@gmail.com}
\affiliation{
   Blackett Laboratory, Imperial College London, London SW7 2AZ, United Kingdom
   }   
\author{Derek K.~K.~Lee}
\affiliation{
   Blackett Laboratory, Imperial College London, London SW7 2AZ, United Kingdom
   }
\author{Johannes Knolle}
\affiliation{
   Blackett Laboratory, Imperial College London, London SW7 2AZ, United Kingdom
   }
\date{\today}
\begin{abstract}
We study the effects of disorder on a Kitaev chain with longer-range hopping and pairing terms which is capable of forming \textit{local} zero energy excitations and, hence, serves as a minimal model for localization-protected edge qubits. The clean phase diagram hosts regions with 0, 1, and 2 Majorana zero-modes (MZMs) per edge. Using a semi-analytic approach corroborated by numerical calculations of the entanglement degeneracy, we show how phase boundaries evolve under the influence of disorder. While in general the 2 MZM region is stable with respect to moderate disorder, stronger values drive transition towards the topologically trivial phase. We uncover regions where the addition of disorder \textit{induces} local zero-modes absent for the corresponding clean system. Interestingly, we discover that disorder destroys any direct transition between phases with zero and two MZMs by creating a tricritical point at the 2-0 MZM boundary of the clean system. Finally, motivated by recent experiments, we calculate the characteristic  signatures of the disorder phase diagram as measured in dynamical local and non-local ``qubit" correlation functions. Our work provides a minimal starting point to investigate the coherence properties of local qubits in the presence of disorder.
\end{abstract}

\maketitle

The Kitaev superconductor $p$-wave chain \cite{kitaev2001} supports an inherently non-local ground state degeneracy, making it possible to create a qubit which is highly fault-tolerant to local decoherence processes \cite{alicea2012} interesting for quantum information technology \cite{alicea2011non}. While the non-locality of the zero-mode offers this protection, it also poses an experimental challenge since non-local measurements are generally difficult \cite{prada2017,jeon2017distinguishing}. Local qubits formed at the edge of topological chains can exhibit remarkable coherence properties when the bulk system is in a many-body localized state \cite{bahri2015,kemp2017}. Intuitively, this can be understood from the localization of bulk excitations, which once excited thermally for example, typically decohere the qubit state. However, their detrimental effects may be suppressed in disordered systems due to lack of thermalization \cite{bahri2015,else2017}. This potentially provides a route to achieve both stable qubits and better experimental controllability.

Motivated by these ideas, we study the effects of disorder on the topological phase diagram for a minimal variant of the Kitaev chain which enables both local and non-local qubit formation \cite{niu2012}. (This system can also be viewed as a three-spin 'cluster' transverse-field Ising model via Jordan-Wigner transformation \cite{smacchia2011}.) The $\mathbb{Z}$ classification \cite{ryu2010} of the model allows two Majorana zero-modes (MZMs) to form on each side of the chain which can pair up to form a \textit{local} zero-energy excitation \footnote{A comment on notation: we use 'zero-mode' to mean zero energy excitation, which pairs 2 MZMs to form a complex fermionic Bogoliubov quasiparticle.}.  
Previous  works have investigated the role of disorder in topological chains with MZMs \cite{gergs2016,degottardi2013,cai2013,lobos2012,degottardi2011,altland2014, mondragon2014,mcginley2017} and found that the clean phase diagram is generally robust for weak values of disorder but strong disorder drives the system into a topologically trivial phase.  While the issue is of central importance in conventional solid-state setups \cite{mourik2012,das2012,deng2016}, recent experiments have devised ``controllable disorder" in the realm of ultracold atoms which allows for precise trajectories in parameter space \cite{meier2018}. Somewhat counter-intuitively, the localizing properties of moderate disorder can also facilitate edge mode formation and induce a topological phase in certain models \cite{li2009, mcginley2017,gergs2016}.

\begin{figure}[b]
\begin{centering}
\includegraphics[scale=0.4]{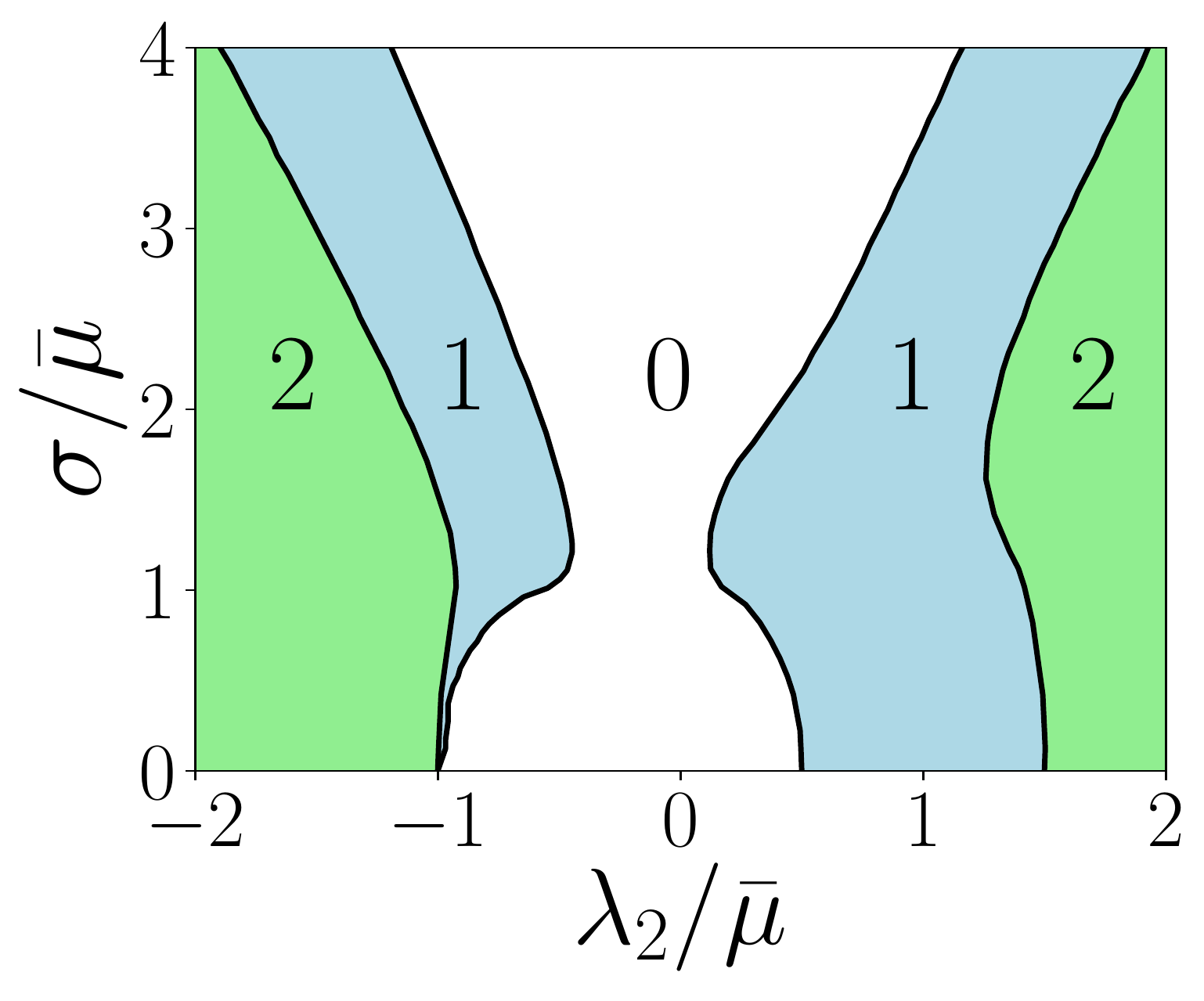}
\par\end{centering}
\caption{\label{fig:lyaPD}Phase diagram indicating the number of MZMs per edge as a function of disorder strength $\sigma$ for $ \lambda_1 / \bar{\mu}=0.5$. The diagram is appears qualitatively similar for any choice of $\lambda_1 / \bar{\mu} \in [0, 2]$. Note the ``tricritical'' point $\lambda_2/\bar\mu=-1$ on the clean axis where all three phases meet. Lyapunov exponents are calculated for chains of  $10^6$ sites. }
\end{figure}

In this work, we show that disorder can induce topological phases with and without local zero-modes in the longer-range Kitaev chain
with equal hopping/pairing for nearest and next-nearest neighbors \cite{niu2012}, described by the Hamiltonian: 
\begin{align}\label{eq:ham}
H= 2&\sum_{i=1}^{N} \mu_{i} a_{i}^{\dagger}a_{i}-\lambda_{1}\sum_{i=1}^{N-1}\left(a_{i}^{\dagger}a_{i+1}+a_{i}^{\dagger}a_{i+1}^{\dagger}+h.c.\right)\notag\\
&-\lambda_{2}\sum_{i=2}^{N-1}\left(a_{i-1}^{\dagger}a_{i+1}+a_{i-1}^{\dagger}a_{i+1}^{\dagger}+h.c.\right)
\end{align}
where $a_i$ are annihilation operators of spinless complex fermions corresponding to site $i$ in the chain, $\lambda_1,\lambda_2,\mu_i \in \mathbb{R}$, and $\mu_i=\mu$ in the clean case.
The disorder is introduced in the on-site energy $\mu_{i}=\bar{\mu}+X(-\sigma,\sigma)$ where $X$ is a uniform random variable chosen between $-\sigma$ and $+\sigma$.

Using a semi-analytical approach relying on transfer matrices and entanglement metrics, we identify a rich disorder phase diagram. This is summarized in Fig.~\ref{fig:lyaPD} which shows phases with 0, 1, or 2 MZMs. It suggests that there exists intermediate regions where local zero-modes remain topologically protected while bulk modes are highly localized.
Indeed, we find examples of induced 1-2 MZM transitions with local qubit formation upon adding disorder, as well as conventional ``topological Anderson insulator" behavior \cite{li2009}.

In light of recent experiments  \cite{meier2018}, we also demonstrate that certain dynamical ``qubit" correlation functions exhibit phase-sensitive signatures even in the presence of disorder \cite{gomez2018,gomez2016,heyl2018detecting}. Specifically, we find that the appearance of (non)-local zero-modes will manifest themselves in the long-time behavior of (non)-local correlators. This provides an experimentally useful diagnostic to detect disorder-induced transitions.

\textit{The clean system.}
We begin our discussion by reviewing the clean phase diagram of the model using transfer matrices.
It is  convenient to define Majorana modes: $\alpha_{2i}=a_{i}^{\dagger}+a_{i},\alpha_{2i+1}=-i(a_{i}^{\dagger}-a_{i})$. The Hamiltonian transforms to a next-nearest-neighbor Su-Schrieffer-Heeger model \cite{su1979} where the sublattice label corresponds to the two Majorana states in the original problem.

\begin{figure}
\begin{centering}
\includegraphics[scale=0.35]{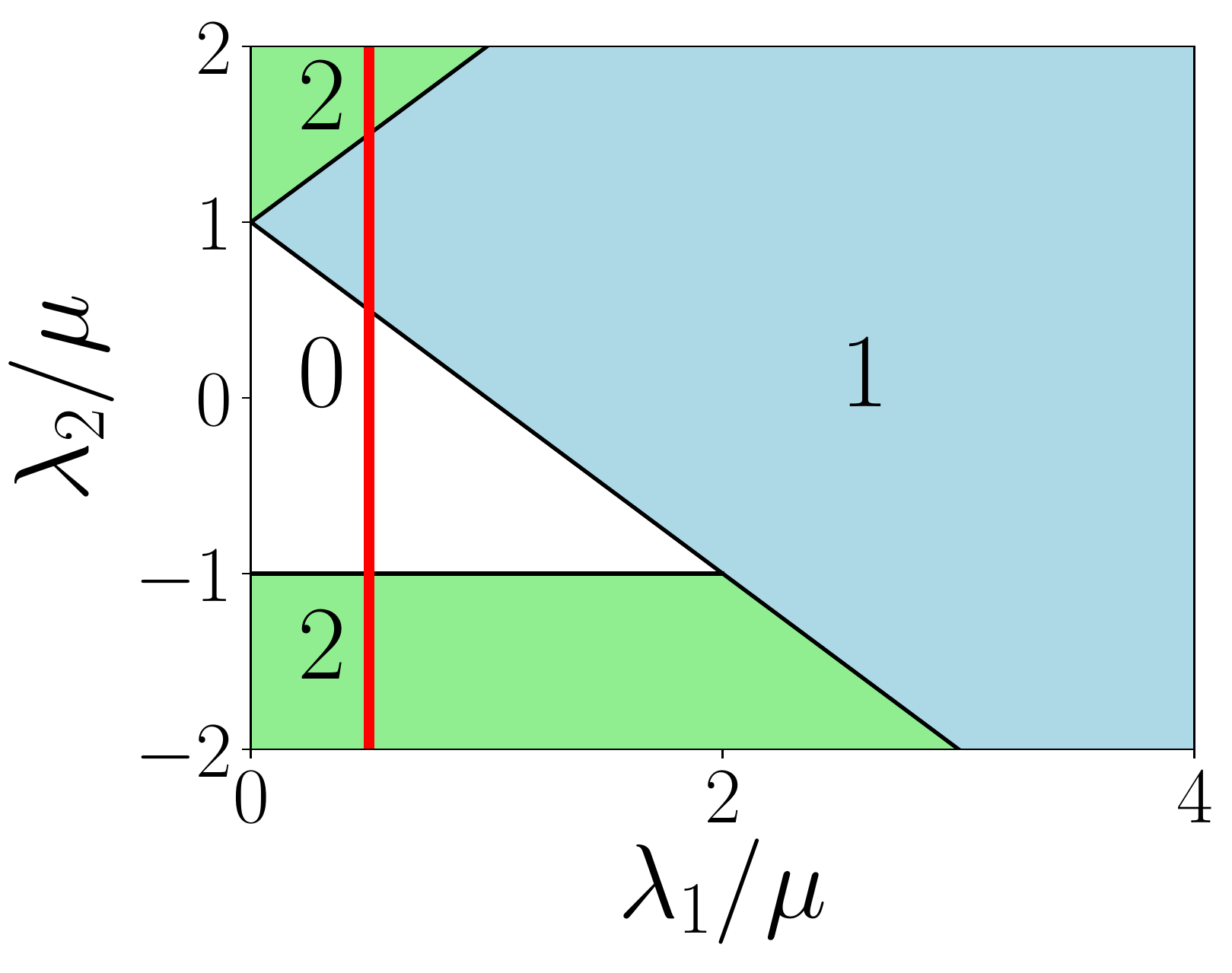}
\par\end{centering}
\caption{\label{fig:Clean-phase-diagram}Clean phase diagram indicating number of MZMs per edge obtained by considering eigenvalues of the transfer matrix $S$. The same diagram can be obtained by finding the winding number of the Bloch Hamiltonian \cite{niu2012}. Red line: clean limit of Fig.~\ref{fig:lyaPD}.}
\end{figure}

To check for the presence of zero-energy edge modes with support on ``odd" sublattice sites on the left edge, we use the eigenvector ansatz $e_{0}=\sum_{i=1}^N A_i \alpha_{2i-1}$. Imposing $H e_{0}=0$ in a semi-infinite geometry, we arrive at the recursion relation
\begin{equation}
\left(\begin{array}{cc}
0 & 1\\
\frac{\mu}{\lambda_{2}} & -\frac{\lambda_{1}}{\lambda_{2}}
\end{array}\right)\left(\begin{array}{c}
A_{i}\\
A_{i+1}
\end{array}\right)=\left(\begin{array}{c}
A_{i+1}\\
A_{i+2}
\end{array}\right).
\end{equation}
Defining this equation as $S\boldsymbol{A}_{i}=\boldsymbol{A}_{i+1}$,  we find $\boldsymbol{A}_{n+1}=S^{n}\boldsymbol{A}_{1}$. The eigenvalues of the transfer matrix $S$ are
\begin{equation}
S\boldsymbol{y}^{\pm}=z^{\pm}\boldsymbol{y}^{\pm},\qquad z^{\pm}=\frac{-\lambda_1 \pm \sqrt{4\lambda_{2}\mu+\lambda_{1}^{2}}}{2\lambda_{2}}.\label{eq:evals}
\end{equation}
These candidate zero-energy edge modes must be normalizable which requires $\left|z\right|<1$. Then, $1/|z|$ gives the localization length of the edge mode.
We confirm in Fig.~\ref{fig:Clean-phase-diagram} the phase diagram found in Ref.~\cite{niu2012} by checking this condition for the  two eigenvalues $z$ as a function of $\lambda_{1},\lambda_{2}$ and $\mu$.
We see that there are phases with 2 MZMs for $|\lambda_{2}/ \mu|>1$.
We note that, when $\lambda_{2}/ \mu<-1$, the two MZMs can have the same localization length since $\sqrt{4\lambda_{2}\mu+\lambda_{1}^{2}}$ can be purely imaginary. This is why two MZMs appear together as we cross the phase boundary at $\lambda_{2}/ \mu=-1$ giving a direct 2-0 MZM transition.

\textit{Disorder Phase Diagram.}
Having found the phase diagram for the clean case by considering the normalization of the edge modes via transfer matrices, a natural question is whether the same can be done for the disordered case. In what follows, we will disorder the chemical potential at each site $\mu_{i}=\bar{\mu}+X(-\sigma,\sigma)$ where $X$ is a uniform random variable chosen between $-\sigma$ and $+\sigma$. In the disordered case, the transfer matrix $S$ now acquires a site dependence $S_{i}$, leading to
\begin{equation}
\boldsymbol{A}_{n+1}=Q_{n}\boldsymbol{A}_{1},\qquad Q_{n}\equiv\prod_{i=1}^{n}S_{i}.
\end{equation}
We are ultimately interested in the large-$n$ behavior of the eigenvalues  $\zeta_{n}^{(1,2)}$ of $Q_{n}$, characterized by
the two ``Lyapunov exponents'' (LEs) $\gamma_{1,2}$:
\begin{equation}
\left|\zeta_{n}^{(1,2)}\right| \propto e^{\gamma_{1,2}n}, \qquad n \gg 1.
\end{equation}
The number of negative LEs will tell us how many normalizable MZMs are present at a given value of $\lambda_{1},\lambda_{2},\bar{\mu},\sigma$ and we can construct phase diagrams accordingly, in direct analogy with the clean case. The construction of a phase diagram is thus equivalent to solving for LEs, which we find using a combination of numerical and analytical techniques.

First, we demonstrate how to find the sum of LEs analytically. The determinant is the product of eigenvalues:
\begin{equation}
\left|\det\left(Q_{n}\right)\right| \propto e^{(\gamma_{1}+\gamma_{2})n}, \qquad n \gg 1.
\end{equation}
The determinant product rule
$\det\left(Q_{n}\right)=\prod_{i=1}^{n}\det\left(S_{i}\right)$
allows us to treat the right-hand-side as a product of random variables. Using the law of large numbers \cite{crisanti1993} leads to
\begin{equation}
\gamma_{1}+\gamma_{2} =\!\!\int \! p(\mu)\ln\left|\det S(\mu)\right|d\mu=\frac{1}{2\sigma}\int_{\bar{\mu}-\sigma}^{\bar{\mu}+\sigma}\!\!\!\!\ln\left|\frac{\mu}{\lambda_{2}}\right|d\mu
\label{eq:gamSumEq}
\end{equation}
for a uniform distribution for the random variable $\mu$.
If $\gamma_{1}+\gamma_{2}<0$, there exists at least one MZM. On the other hand, $\gamma_{1}+\gamma_{2}>0$ means that there can be at most one MZM. This separates parameter space into two regions which interestingly do not depend on $\lambda_{1}$.

Second, we still need one more constraint in order to determine both LEs for a point in parameter space. While analytical solutions for the exponents are rare, it is straightforward to calculate LEs numerically for each realization \cite{crisanti1993,castanier1995,kramer1993}. We use the Wolf algorithm \cite{wolf1985} to determine  the maximum LE, which is discussed in Supplementary Material 1.

Having found the maximum and sum of Lyapunov exponents, we can construct the phase diagram (Fig.~\ref{fig:lyaPD}). We notice that, for weak disorder, the diagram is generally unaltered from the clean case (with the notable exception near $\lambda_2/ \bar{\mu}=-1$). Physically, this implies that a local qubit constructed out of two MZMs on one side of the chain  is able to withstand moderate amounts of disorder. Nevertheless, sufficiently strong disorder induces 2-1 and 1-0 MZM transitions. This can occur because two MZMs do not necessarily possess the same localization length, hence disorder can push one mode into the bulk whilst the other remains. Remarkably, we also find the converse situation in other parts of the phase diagram. Some parts of the clean 0-MZM phase can support an edge mode when disorder is added. Similarly, a 1-MZM phase can support 2 MZMs upon addition of disorder (near $\lambda_2 / \bar{\mu}=1.5$). This is an analog of the disorder-induced topological Anderson insulator for  $\mathbb{Z}$-classified models \cite{li2009}.

\begin{figure}
\begin{centering}
\includegraphics[scale=0.35]{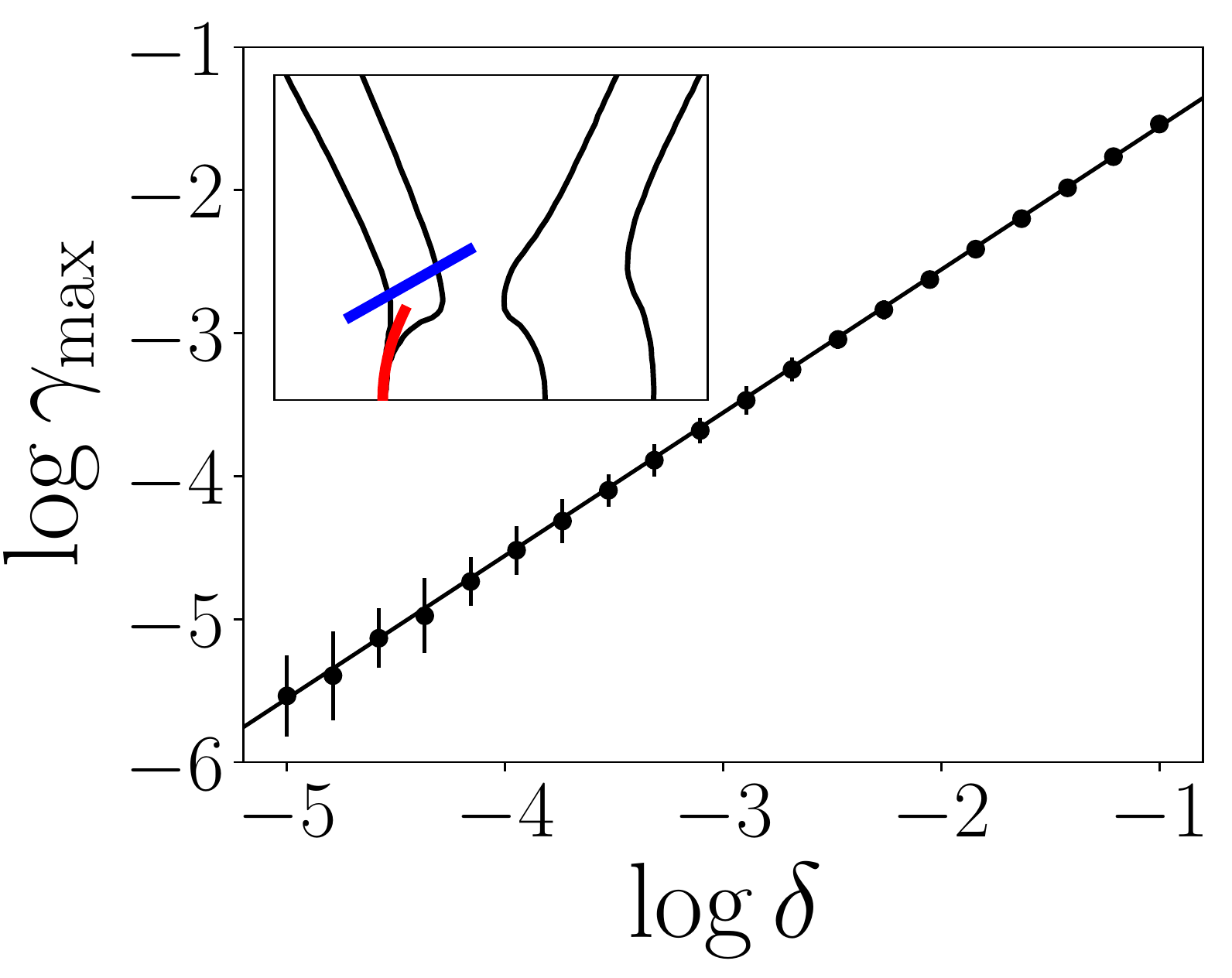}
\par\end{centering}
\caption{\label{fig:logPlot} The maximum Lyapunov exponent as a function of $\delta$ (which parameterizes the $\gamma_{1}+\gamma_{2}=0$ path, depicted in red in the inset) on a log-log plot over four decades. ($ \lambda_1 / \bar{\mu}=0.5$.) The fitted slope of 1 implies $\gamma_{\max}\propto\delta\propto\sigma^{2}$ along this path. We use a chain with $2\times 10^{7}$ sites and average over 10 configurations to estimate errors which increase when the localization length becomes comparable to the length of the chain.
}
\end{figure}

Let us now look more closely at the system near $\lambda_2/ \bar{\mu}=-1$. Recall that the clean localization length near the 2-0 boundary is the same for both modes. Disorder is responsible for a splitting of the Lyapunov exponents $\gamma_{1,2}$ of the transfer matrix $Q_n$ in this region. This leads to an anomalous ``tricritical point" at $\lambda_2/ \bar{\mu}=-1, \sigma=0$ which separates 2, 1, and 0 MZM regions. We can understand this behavior if we focus on the line in parameter space where $\gamma_1=-\gamma_2$, which we can determine analytically according to Eq.~(\ref{eq:gamSumEq}). The line is either an exact phase boundary (if $\gamma_1=\gamma_2=0$), or it must lie in the 1-MZM region due to the splitting. At weak disorder, this line is given by $\sigma/ \bar{\mu}=\sqrt{6 \delta}$ where $\delta=\lambda_2/ \bar{\mu}+1$ (red line in Fig.~\ref{fig:logPlot} inset). In Fig.~\ref{fig:logPlot}, we present numerical evidence that $\gamma_\text{max} \propto \delta \propto \sigma^2$. This suggests that  a 1-MZM phase arises between the 2- and 0-MZM phases for an arbitrary amount of disorder. Indeed, this behavior is generic for all $\lambda_1/\bar{\mu}$.

\begin{figure}
\begin{centering}
\includegraphics[scale=0.35]{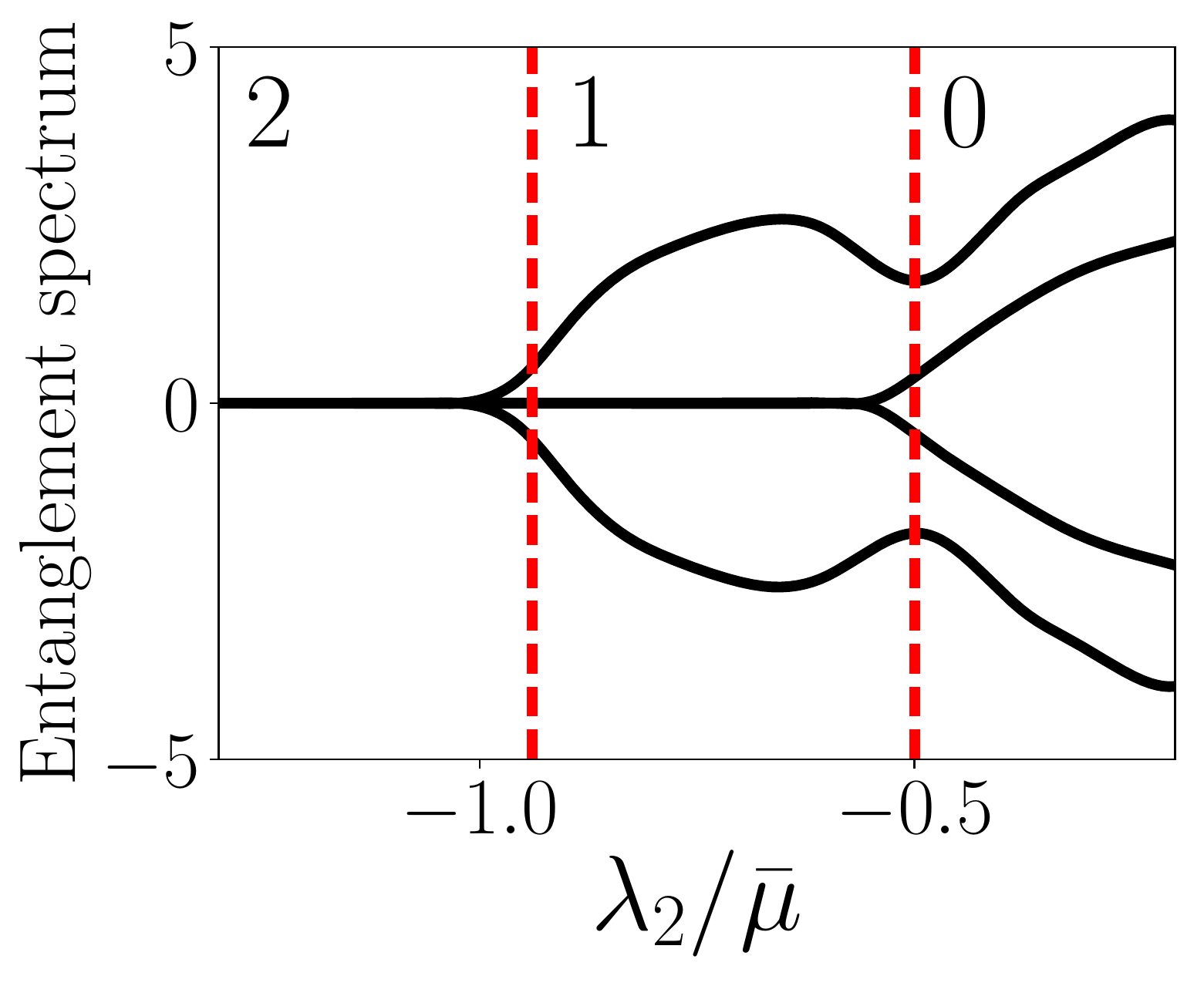}
\par\end{centering}
\caption{\label{fig:entDeg} Four modes closest to zero energy of the entanglement spectrum (defined in the text) along the line $\sigma = 0.75 \lambda_2 +2 \bar{\mu}$, $ \lambda_1 / \bar{\mu}=0.5$ (blue line in Fig.~\ref{fig:logPlot} inset). Red dashed lines indicate the phase transitions based on Lyapunov exponents. The spectrum has 4, 2, or 0 modes with zero energy depending on the topological phase of the system. For the simulation we use $N=1000$ particles, disorder averaged across 1000 configurations.}
\end{figure}

\begin{figure*}
\begin{centering}
\includegraphics[scale=0.35]{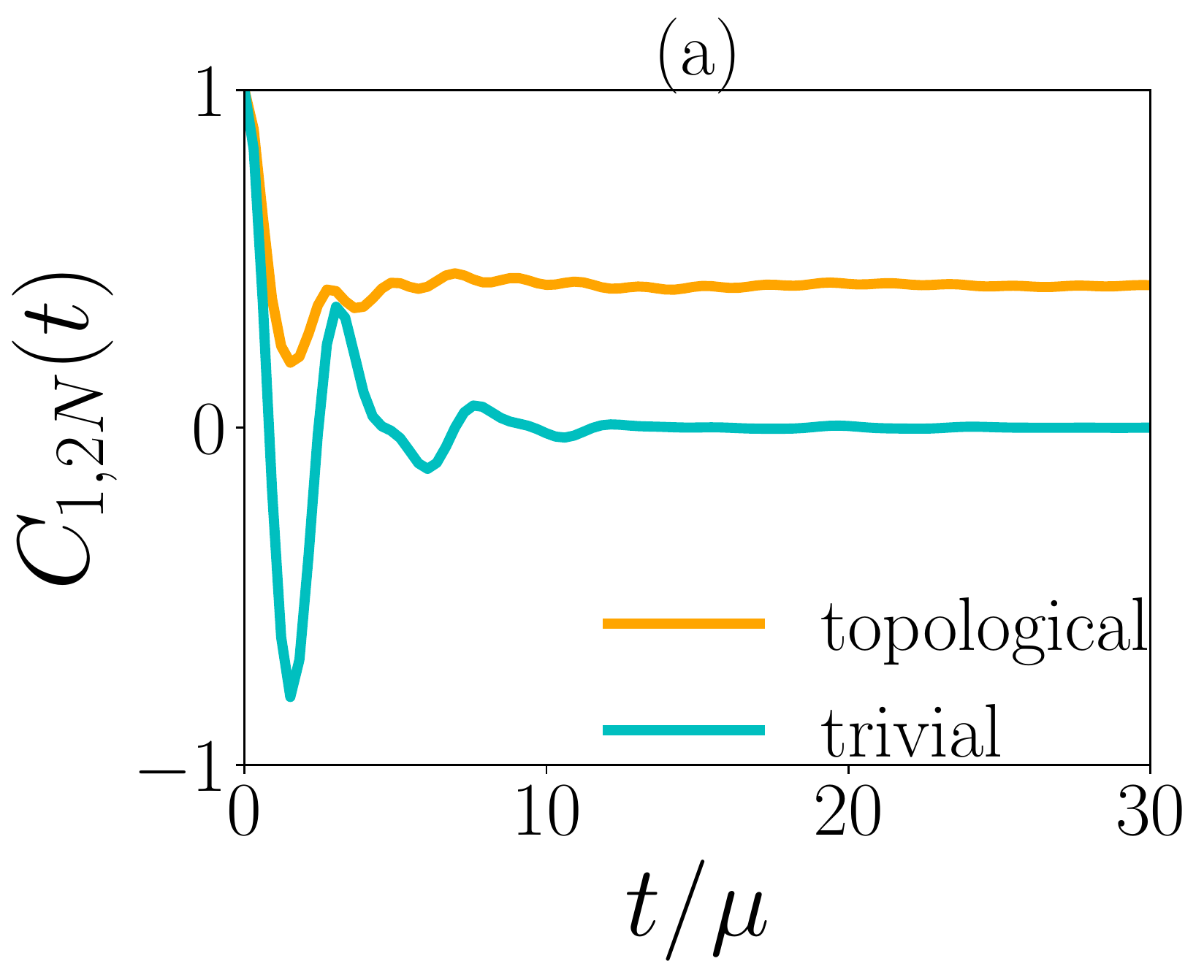}
\includegraphics[scale=0.35]{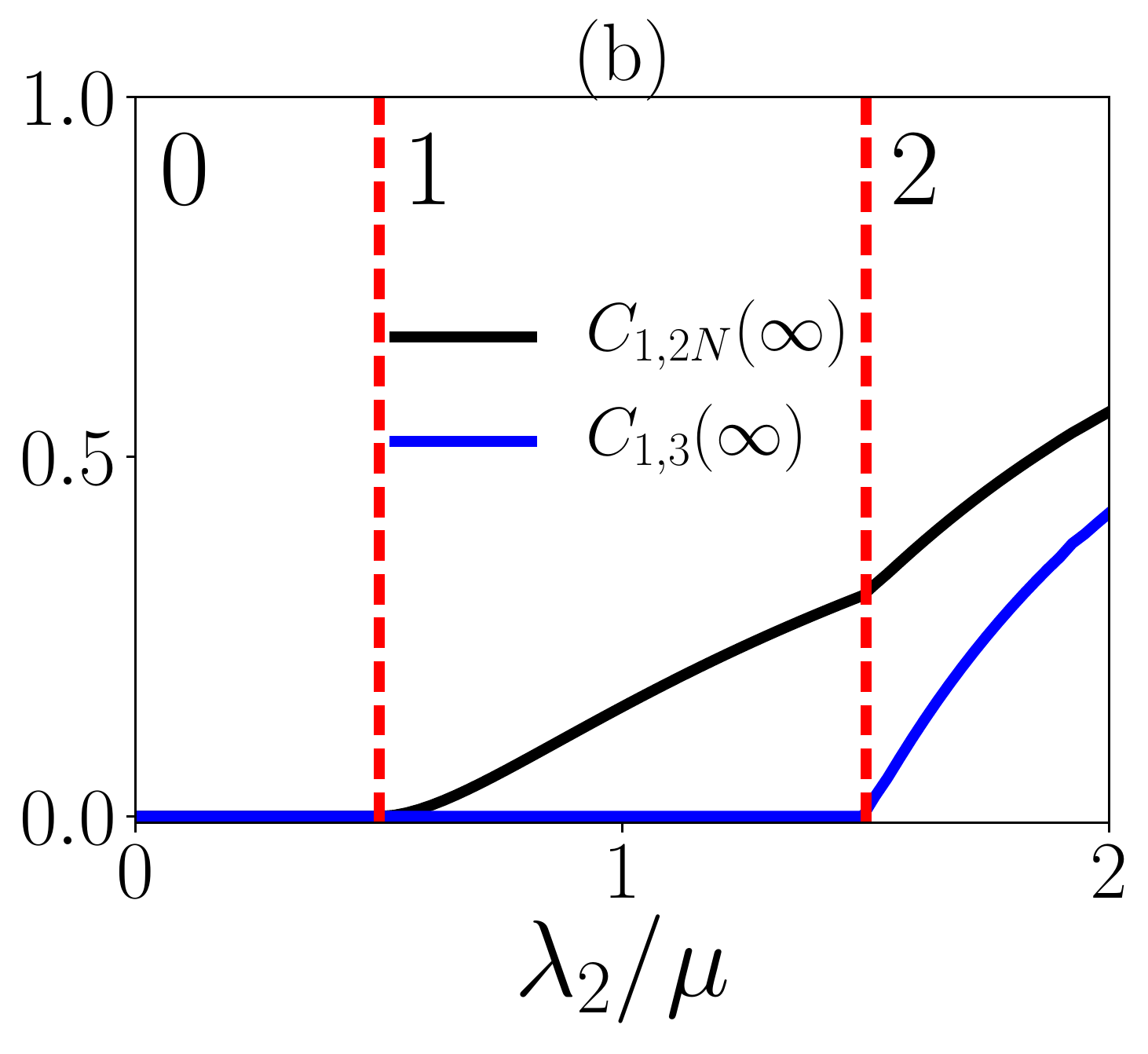}
\includegraphics[scale=0.35]{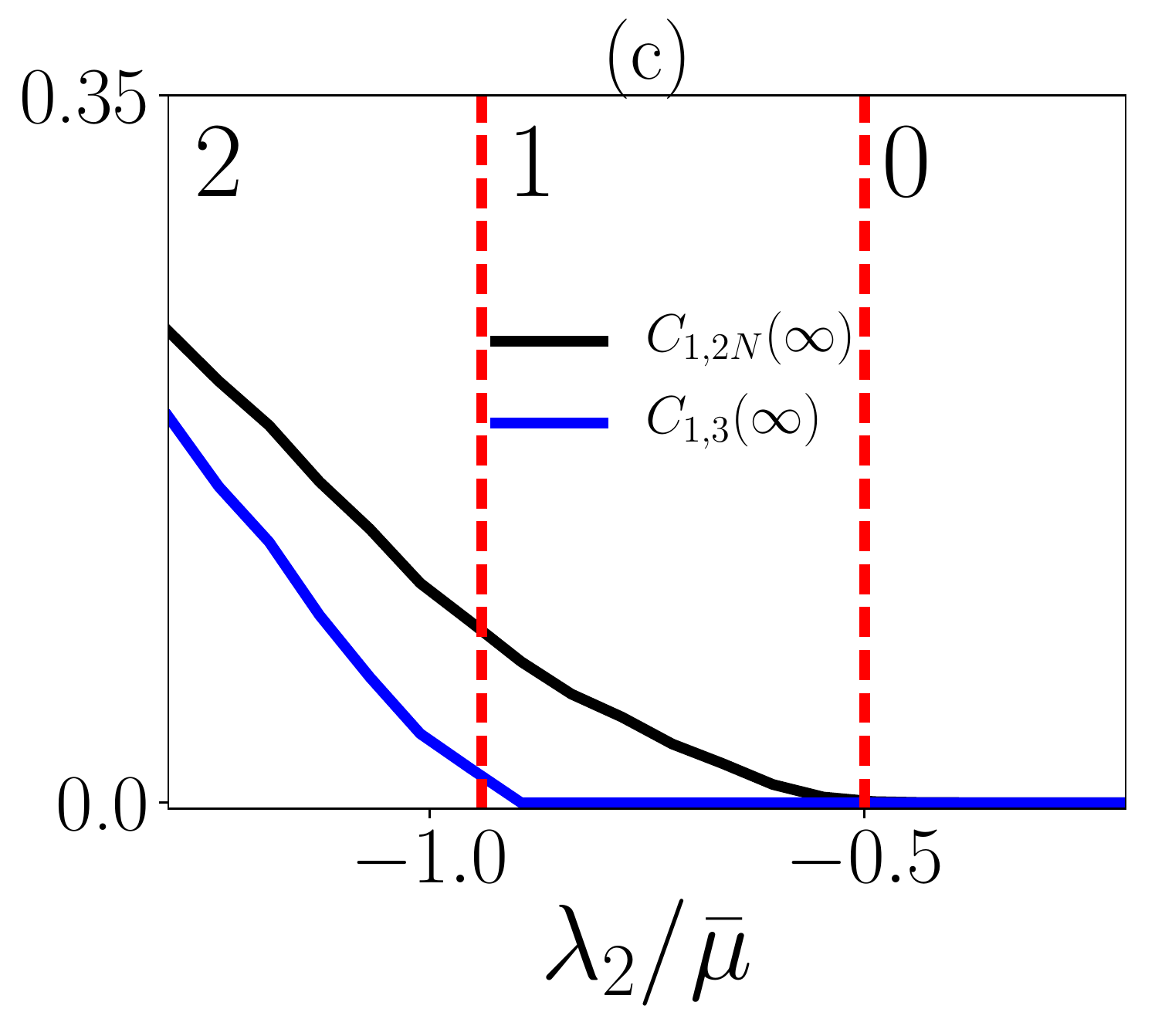}
\par\end{centering}
\caption{\label{fig:op} (a) Time-dependent, non-local correlator for clean systems at $\lambda_1/\mu = 0.5$, $\lambda_2/\mu = 0.45$ (trivial) and 2.0 (topological). If there are no MZMs (trivial), the long-time value is zero, otherwise it saturates at a non-zero value. (b) Clean long-time saturation value when $\lambda_1/\mu=0.5$. The non-local (local) correlator signals the onset of the 1 (2)-MZM phase. (c) Disorder-averaged long-time saturation value along the line $\sigma = 0.75 \lambda_2 +2 \bar{\mu}$ when $\lambda_1/\bar{\mu}=0.5$ over $10^4$ samples. Behavior is analogous to the clean case. Red dashed lines indicate the phase transitions based on Lyapunov exponents.}
\end{figure*}

In a third step, we use the ``entanglement degeneracy criterion" to validate our phase diagram \cite{li2008,turner2011,yates2017, gergs2016}. More specifically, we consider the chain in a Majorana basis, impose periodic boundary conditions, and partition the system into two halves. The entanglement spectrum is defined as the set of eigenvalues of the operator $\bar{H}=-\log \text{tr}_{N} \rho_{gs}$ where $\rho_{gs}$ is the ground state density matrix and we trace over the Majorana modes $\alpha_i$ on half the chain ($1\le i\le N$). The degeneracy of the entanglement spectrum can be used to distinguish distinct topological phases, since the number of entanglement zero-modes coincides with the number of topologically-protected MZMs in the finite system. In Fig. \ref{fig:entDeg}, we plot the four modes closest to zero ``energy" in the entanglement spectrum. Indeed we find that gaps in the spectrum align with the phase boundaries derived via Lyapunov exponents. We refer to Supplementary Material 2 with Refs. \cite{peschel2003,peschel2009, latorre2003} for details concerning the calculation.

\textit{Signatures of Dynamical Correlations.} 
We demonstrate that certain time-dependent, \textit{local} correlation functions can identify the onset of a 2-MZM phase, while \textit{non-local} correlators identify the 1 MZM phase. Recent studies \cite{gomez2018,heyl2018detecting} have found that the long-time values of certain ``qubit" correlators will saturate in a topological phase, while they tend to zero in the trivial phase, reminiscent of order parameter behavior. Consider the correlator
\begin{equation}
C_{i,j}(t)=\frac{1}{2}\left\langle \text{vac}\left|\left\{ e^{iHt} q_{i,j} e^{-iHt}, q_{i,j}\right\} \right|\text{vac}\right\rangle 
\end{equation}
where $q_{i,j}=i\alpha_{i}\alpha_{j}$ and $\left|\text{vac}\right\rangle $ is the vacuum of Bogoliubov quasiparticles. Note that $q_{i,j}$ is both Hermitian and unitary which restricts its eigenvalues to $\pm 1$. We find
\begin{align}\label{eq:corrs}
    &C_{i,j}(t)=\left|\sum_{m=1}^{N}M_{i,m}M_{j,N+m}\right|^{2}+\notag\\ &\ \sum_{\stackrel{m=1}{n>m}}^{N}\!
    \cos\left[\left(\epsilon_{m}+\epsilon_{n}\right)t\right]\left|M_{i,N+m}M_{j,N+n}-(m \leftrightarrow n)\right|^{2}
\end{align}
where $M= P^{-1} U$, $U$ is a unitary transformation which diagonalizes the Hamiltonian (\ref{eq:ham}) written in the fermionic basis ($a,a^\dagger$), $\boldsymbol{a}=P\boldsymbol{\alpha}$,  and $\epsilon_i$ is the quasiparticle energy of the $i$th mode (see Supplementary Material 3 for details). The long-time behavior $t \rightarrow \infty$ is dominated by the first term and pairs of zero energy modes $m',n'$ such that $\cos\left[\left(\epsilon_{m'}+\epsilon_{n'}\right)t\right]=1,  \forall t$.

In Fig.~\ref{fig:op}(a), we show that the long-time behavior of the non-local correlator between the end sites will saturate to a non-zero value in a topological phase ($>$0 MZMs), whilst the correlator tends to zero in a 0-MZM phase. In Fig.~\ref{fig:op}(b) we plot the long-time saturation of the non-local correlator $C_{1,2N}$ as well as the local correlator $C_{1,3}$ for the clean system. We find that the former serves as an indicator of the 0-1 MZM transition, while the latter signals the 1-2 MZM transition. This is in agreement with our understanding that we require two MZMs with support on odd ``sublattice" sites in order to create a local qubit state.

In Fig. \ref{fig:op}(c), we plot the disorder-averaged correlators across a 2-1-0 path (the same one taken in Fig. \ref{fig:entDeg}) and find analogous behavior to the clean case. We point out that the standard deviation is quite high at non-zero saturation ($\sim 0.1$) indicating a strong configurational dependence. Nevertheless, one would anticipate a disordered experiment to exhibit non-zero long-time saturation which ought to be observable. We have thus found another useful metric to describe and experimentally diagnose topological phase transitions in the presence of disorder.

\textit{Discussion.}
In summary, we have investigated the disorder phase diagram of a minimal model which supports two MZMs on each edge of the chain, thus capable of forming a local zero energy excitation representing a qubit. We discover regions of moderate disorder where 2 MZMs are stable, while strong disorder generally drives a 2$\rightarrow$1$\rightarrow$0 MZM transition. Physically this can occur because two MZMs with the same clean localization length can split in the presence of disorder, with one of them vanishing into the bulk before the other. In addition, we have discovered regions where the addition of disorder can induce local zero-modes. We have found a special tricritical point where the 2, 1, and 0-MZM regions meet as corroborated by a combination of analytical and numerical tools. 

From a methodological point of view, we have demonstrated that the calculation of Lyapunov exponents is a computationally cheap and intuitive way of obtaining phase diagrams which agree with conventional methods such as entanglement degeneracy for $\mathbb{Z}$-classified models. Moreover, we have argued that signatures of the transition ought to be observable in the measurement of both non-local and local correlation functions, the latter of which might be beneficial to experimental setups.

While the Hamiltonian (\ref{eq:ham}) describes a mean-field superconductor with disordered chemical potential, it can also be viewed as an exact three-spin Ising model with a disordered transverse field via Jordan-Wigner transformation \cite{niu2012}. Renormalization group analysis naturally generates three-spin terms in the 1D transverse field Ising model \cite{hirsch1979} which indicates that Eq.~(\ref{eq:ham}) appears as an effective low-energy description. In addition, our work opens the possibility of using disorder-induced local zero-modes as protected qubits  \cite{bahri2015}. We have shed light on the critical upper bound of disorder beyond which we no longer expect topological MZMs to be protected, and how their presence can be diagnosed in local dynamical correlation functions. Future work should aim to explore the stability and coherence of local qubits away from the zero-temperature, non-interacting, and closed system limit in the presence of quenched disorder. 

\textit{Acknowledgements.}
We would like to thank Angus MacKinnon and Caleb Q. Cook for useful discussions.  S.L. is supported by the Imperial College President's Scholarship.

\bibliography{disorder} 
\bibliographystyle{apsrev4-1}
\newpage
\widetext

\setcounter{equation}{0}
\setcounter{figure}{0}
\setcounter{table}{0}
\setcounter{page}{1}
\makeatletter
\renewcommand{\theequation}{S\arabic{equation}}
\renewcommand{\thefigure}{S\arabic{figure}}

\subsection{\large Supplemental Material 1: Maximum Lyapunov Exponent}
\label{supp:wolf}

Here, we present the Wolf algorithm \cite{wolf1985} which we use to find the maximum Lyapunov exponent. Consider the $2 \times 2$ random matrix product $Q_{n}\equiv\prod_{i=1}^{n}S_{i}$. The eigenvalues $\zeta_{n}^{(1,2)}$ of $Q_{n}$ scale as
\begin{equation}
  \left|\zeta_{n}^{(1,2)}\right| \propto e^{\gamma_{1,2}n}\,,\qquad n\gg 1.
\end{equation}
We wish to find the bigger of the two Lyapunov exponents, which we refer to as $\gamma_{\text{max}}$.
\begin{enumerate}
 \item Begin with an arbitrary vector $\boldsymbol{r}_{0}$ which is normalized to 1 and multiply it by $m$ iterations of the random matrix $S_{i}$ from the left.
\item Store the norm of the resulting vector as $R_{0}$ and renormalize the vector to 1. Generically, at step $k$ we store $R_{k}=\left|\boldsymbol{r}_{k}'\right|,\boldsymbol{r}_{k}'\equiv\left(\prod_{j=km+1}^{(k+1)m}S_{i}\right)\boldsymbol{r}_{k}$ and renormalize the resulting vector $\boldsymbol{r}_{k+1}\rightarrow\boldsymbol{r}_{k}'/\left|\boldsymbol{r}_{k}'\right|$.
\item Repeat the procedure $k=N$ times.
\item The maximum Lyapunov exponent is given by $\displaystyle{\gamma_{\text{max}}=\lim_{N\rightarrow\infty}\frac{1}{(N+1)m}\sum_{k=0}^{N}\ln\left[R_{k}\right]}$.
\end{enumerate}
This procedure relies on the fact that the norm of an arbitrary vector will grow according to the largest eigenvalue of the random matrix product, in analogy with a deterministic product. We split up the product into bins of size $m=20$ in order to avoid numerical error, since successive multiplication of the random matrix $S_i$ will generically cause the norm of $\boldsymbol{r}$ to tend to either zero or infinity (away from a phase boundary).

\subsection{\large Supplemental Material 2: Entanglement Spectrum}
\label{supp:ent}

In this section we provide some details on how to calculate the entanglement spectrum for the model considered, following closely to Refs. \cite{mcginley2017, latorre2003}. Reference \cite {peschel2009} demonstrates that for a quadratic Majorana Hamiltonian,  the entanglement spectrum is related to the eigenvalues of the ground state subsystem correlation matrix via
\begin{equation}
\tanh \left( \frac{e_{\text{ent}}}{2}\right) = e_{\text{corr}} - 1
\end{equation}
where $e_{\text{ent}}$ are the eigenvalues of the ``entanglement Hamiltonian" $\bar{H} = -\log \text{tr}_{N} \rho_{\text{gs}}$, and $e_{\text{corr}}$ are the eigenvalues of the ground state subsystem correlation matrix $\kappa_{\text{sub}}$. In our simulation Fig. \ref{fig:entDeg}, we calculate the entanglement spectrum by tracing over half the degrees of freedom in the Majorana basis $\alpha$. Hence we will show how to find the full correlation matrix in a Majorana basis: $\kappa_{ij} =\left\langle \text{vac}\left|\alpha_{i}\alpha_{j}\right|\text{vac}\right\rangle$, from which one can read off the subsystem correlation matrix by considering all degrees of freedom which have not been traced away.

The Hamiltonian reads
\begin{equation}
H=\boldsymbol{a}^{\dagger}\tilde{H}\boldsymbol{a},\qquad\boldsymbol{a}=\left(a_{1},\ldots,a_{N},a_{1}^{\dagger},\ldots,a_{N}^{\dagger}\right)^{T}
\end{equation}
where $a_{i}^{\dagger}$ is the fermion creation operator at lattice site $i$. The Hamiltonian is diagonalized via a unitary Bogoliubov transformation according to 
\begin{equation}
H=\boldsymbol{b}^{\dagger}\Lambda\boldsymbol{b},\qquad\boldsymbol{a}=U\boldsymbol{b},\qquad U^{\dagger}\tilde{H}U=\Lambda,\qquad\Lambda=\text{Diag}\left[+\epsilon_{1},\ldots,+\epsilon_{N},-\epsilon_{1},\ldots,-\epsilon_{N}\right].
\end{equation}
We can define a Majorana basis for both $b$ and $a$ fermion operators according to the following transformation $\boldsymbol{b}=P\boldsymbol{\beta}$, $\boldsymbol{a}=P\boldsymbol{\alpha}$. Using these transformations, we can relate the Majorana modes according to
\begin{equation}
\boldsymbol{\beta}=P^{-1}\boldsymbol{b}=P^{-1}U^{-1}\boldsymbol{a}=P^{-1}U^{-1}P\boldsymbol{\alpha}\equiv W\boldsymbol{\alpha}
\end{equation}

The correlation matrix of the vacuum is easily determined in the $\boldsymbol{\beta}$ basis. All Bogoliubov annihilation operators $b_{i}$ will destroy the vacuum state. In other words
\begin{align}
\left\langle \text{vac}\left|b_{p}b_{q}^{\dagger}\right|\text{vac}\right\rangle  & =\delta_{p,q} \\
\left\langle \text{vac}\left|b_{p}^{\dagger}b_{q}\right|\text{vac}\right\rangle  & =0\\
\left\langle \text{vac}\left|b_{p}b_{q}\right|\text{vac}\right\rangle  & =0
\end{align}
By translating the $b$ operators into Majorana $\beta$ operators, it follows that
\begin{equation}
T_{pq}\equiv\left\langle \text{vac}\left|\beta_{p}\beta_{q}\right|\text{vac}\right\rangle =\delta_{pq}+i\Gamma_{pq},\qquad\Gamma_{pq}=\mathbb{I}_{N}\otimes\left(\begin{array}{cc}
0 & 1\\
-1 & 0
\end{array}\right)
\end{equation}
So we have determined the correlation matrix in the $\beta$ Majorana basis.

Finally, we would like to determine the correlation matrix in the $\alpha$ basis. This can be done with the following transformations
\begin{align}
\kappa_{ij} & =\left\langle \text{vac}\left|\alpha_{i}\alpha_{j}\right|\text{vac}\right\rangle \\
 & =\left\langle \text{vac}\left|\left(W^{-1}\boldsymbol{\beta}\right)_{i}\left(W^{-1}\boldsymbol{\beta}\right)_{j}\right|\text{vac}\right\rangle \\
 & =\sum_{pq}W_{ip}^{-1}W_{jq}^{-1}\left\langle \text{vac}\left|\beta_{p}\beta_{q}\right|\text{vac}\right\rangle \\
\Rightarrow \kappa & =W^{-1} T \left(W^{-1}\right)^{T}
\end{align}

\subsection{\large Supplemental Material 3: Dynamical Correlation Functions}
\label{supp:corr}

In this section we will derive Eq. (\ref{eq:corrs}) from the main text. We are interested in the time-dependent correlator
\begin{equation}
C_{i,j}(t)=\frac{1}{2}\left\langle \text{vac}\left|\left\{ e^{iHt} q_{i,j}e^{-iHt},q_{i,j}\right\} \right|\text{vac}\right\rangle 
\end{equation}
where $q_{i,j}=i\alpha_{i}\alpha_{j}$ and $\left|\text{vac}\right\rangle $ is the vacuum of Bogoliubov quasiparticles.

We use the same definitions as Supplementary Material 2. We can relate the Bogoliubov quasiparticles to Majorana modes according to
\begin{equation}
\boldsymbol{b}=U^{-1}\boldsymbol{a}=U^{-1}P\boldsymbol{\alpha}\Rightarrow\boldsymbol{\alpha}=P^{-1}U\boldsymbol{b}\,.
\end{equation}
If we define $M\equiv P^{-1}U$ then $\boldsymbol{\alpha}=M\boldsymbol{b}$.

We will now use these expressions to solve for $C_{i,j}(t)$. Explicitly
\begin{align}
C_{i,j}(t) & =\frac{1}{2}\left(\left\langle \text{vac}\left|e^{iHt}q_{i,j}e^{-iHt}q_{i,j}\right|\text{vac}\right\rangle +\left\langle \text{vac}\left|q_{i,j}e^{iHt}q_{i,j}e^{-iHt}\right|\text{vac}\right\rangle \right)\\
 & =\frac{1}{2}\left(\left\langle \text{vac}\left|q_{i,j}e^{-iHt}q_{i,j}\right|\text{vac}\right\rangle +\left\langle \text{vac}\left|q_{i,j}e^{iHt}q_{i,j}\right|\text{vac}\right\rangle \right)\\
 & =\frac{1}{2}\left(\left\langle \psi_{i,j}\left|e^{-iHt}\right|\psi_{i,j}\right\rangle +h.c.\right),\qquad\left|\psi_{i,j}\right\rangle \equiv q_{i,j}\left|\text{vac}\right\rangle 
\end{align}
We see that $C_{i,j}(t)$ must be real since it is the sum of a complex conjugate pair. Once we note that 
\begin{equation}
\left|\psi_{i,j}\right\rangle =i\left[\sum_{m=1}^{N}\sum_{n>m}^{N}\left(M_{i,N+m}M_{j,N+n}-M_{j,N+m}M_{i,N+n}\right)b_{m}^{\dagger}b_{n}^{\dagger}+\sum_{m=1}^N M_{i,m}M_{j,N+m}\right]\left|\text{vac}\right\rangle 
\end{equation}
we find
\begin{equation}
C_{i,j}(t)=\left|\sum_{m=1}^{N}M_{i,m}M_{j,N+m}\right|^{2}+\sum_{m=1}^{N}\sum_{n>m}^{N}\cos\left[\left(\epsilon_{m}+\epsilon_{n}\right)t\right]\left|M_{i,N+m}M_{j,N+n}-M_{j,N+m}M_{i,N+n}\right|^{2}.
\label{eq:suppCorr}
\end{equation}
which is given in (\ref{eq:corrs}) in the main text.

In the long-time limit $t\rightarrow\infty$ we notice that the terms which dominate $C_{i,j}$ are: 1) the first term in the right-hand-side of (\ref{eq:suppCorr}) which is independent of time, and 2) pairs of zero-energy modes $m',n'$ such that $\cos\left[\left(\epsilon_{m'}+\epsilon_{n'}\right)t\right]=1,  \forall t$. In the case when there is no more than 1 zero-mode, then there are no pairs of zero-modes and the first term is the long-time saturation value. Incidently, the saturation is equivalent to a \textit{static} correlation function 
\begin{equation}
\text{if no pairs of zero-modes:} \lim_{t\rightarrow \infty}C_{i,j}(t)=K_{i,j}\qquad K_{i,j}=\left|\left\langle \text{vac}\right|q_{i,j}\left|\text{vac}\right\rangle \right|^{2}.
\end{equation}
Only in the presence of 2 or more zero-modes does the long-time saturation value differ from this static correlator $K_{i,j}$.
\end{document}